# Influence of Bismuth Oxide as a Sintering Aid on the Densification of Cold Sintering of Zirconia

Bhootpur N, Brouwer H, Tang Y

*Abstract*—In the past decades, Zirconia ($ZrO_2$) has emerged as a promising technical ceramic, both as high temperature structural material and electrolyte for fuel cells, etc. The traditional synthesis of $ZrO_2$ with spark plasma sintering (SPS) usually requires a sintering temperature as high as 1200 °C. General interest in lowering the sintering temperature to reduce energy consumption and thermal stresses has led to research on two promising routes - cold sintering via temperature-dependent chemical reactivity and sintering aids, which facilitates mass transport and improves densification. Here we combine both by developing a single-step sintering process benefitting from both water vapor through the in-situ conversion of $Zr(OH)_4$ to $ZrO_2$ and liquid phase $Bi_2O_3$ as a sintering aid. The resultant $ZrO_2$ has a relative density above 80% with a sintering temperature as low as 900 °C, significantly higher than that of $ZrO_2$ without sintering aids, which had a relative density of 54%, both sintered at 50 MPa. The dependence of porosity of sintered samples as a function of sintering pressure (range: 50MPa - 300MPa) and temperature (range 400 °C – 1200 °C) is mapped out as guidance for further material property design. A linear relationship between hardness and relative density was found, with a maximal hardness of 6.6 GPa achieved in samples with 30% porosity. In addition to sintered density, phase stabilization of tetragonal $ZrO_2$ is enhanced at sintering temperature of 900 °C with water vapor and $Bi_2O_3$, respectively.

## I. Introduction

Zirconium dioxide (Zirconia) is a functional ceramic which could serve both structural and functional purposes in applications such as in aerospace, automotive, dental, medical, and other industries. In the past, numerous studies [1]-[5] on traditional spark plasma sintering of zirconium oxides have discovered that 1200 ºC is the ideal sintering temperature for Zirconia to achieve a relative density of 80 % or higher.

Recent studies [6]-[11] have explored a novel route of sintering ceramics at temperatures lower than their traditional sintering temperatures called the cold sintering technique by utilizing solvents in the sintering system. Thermodynamically, the liquid phase (various polar solvents, such as water) partially dissolves the surfaces of ceramic particulates, followed by their evaporation upon heating, yielding to a supersaturated solution for precipitation; and kinetically, mass transport is facilitated through enhancing the diffusion of solute ions; also, the liquid phase wets the particle surfaces, hence further benefits particle rearrangement for compaction under pressure [7].

Even though cold sintering process has the advantage of reducing energy required in conventional high temperature sintering processes, it also has many disadvantages. Firstly, the densities of the as-sintered samples are often much lower as compared to the traditional sintering protocols. Additionally, the hardness of the cold-sintered samples are also reported to be lesser than the traditionally sintered samples, which affect their application in many industries. As a result, it is not uncommon practice to add a conventional sintering/annealing step to further density the samples, however, this turns into a two-step sintering process, which not only adds complexity but moreover contradicts the original energy-saving purpose.

Hydroxide precursors of Zirconia could convert into $ZrO_2$ and water vapor upon heating and were used as starting powders to achieve low-temperature sintering by Elissalde et al. [12]. A relative density of 70-80% could be achieved after 10 min dwelling at 350 °C and 600 MPa. To promote sintering of ceramic powders with high melting temperatures, sintering aids are often used [13]-[18]. One of the reasons they work is due to liquid phase sintering. The sintering aid melts at a certain temperature and becomes liquid that fills in between the solid powder particles. With a nonzero solubility of solid powder in the liquid phase, the mass transport can be enhanced through diffusion of the solutes and the particle edges can also be wet to facilitate particle rearrangement, both of which is similar to the principles of cold sintering. However, in contrast to cold sintering where the solvent will eventually evaporate and leave the system, sintering aids will remain in the system, most likely in solid solution form. This alters the stoichiometry of the original sintering powders and could affect various properties, such as toughness.

Compared to other various sintering aids, bismuth oxide has a melting temperature at 817 ℃, which is lower than the usual sintering temperatures of $ZrO_2$ (higher than 1200 ℃) but also still high enough to promote decent sintering. Moreover, the solubility of $ZrO_2$ in bismuth oxide was found to be between 2~10 mol % between 800 ℃ and 900 ℃ [19], which is promising to induce liquid phase sintering. While the solubility of $Bi_2O_3$ in $ZrO_2$ is preferably to be small so the liquid phase wouldn't be transient, unfortunately this data remain lacking in literature despite the fact that $Bi_2O_3$ was previously used as a tetragonal phase stabilizer in Zirconia systems at low temperatures by Gulino et al. [20]. One of our goals in this study is thus to investigate whether $Bi_2O_3$ can be an effective liquid phase sintering aid for $ZrO_2$.

In this study, we combine both advantages from cold sintering effect provided by the in-situ conversion of $Zr(OH)_4$ to $ZrO_2$ and $H_2O$ at 400ºC and sintering aided by $Bi_2O_3$ in the

N. Bhootpur is associated with the Delft University of Technology, Delft, The Netherlands NL
H Brouwer, is associated with the Delft University of Technology, Delft, The Netherlands NL

Y. Tang is an assistant professor at the Delft University of Technology, Delft, The Netherlands NL
Y Tang is the corresponding author, e-mail: Y.Tang-5@tudelft.nl).

temperature range of 817 ⁰C - 900 ⁰C, effectively achieving a density of 87 % with a sintering temperature as low as 900 ⁰C in a 'single sintering step'. Different piston pressures (50MPa, 150 MPa, 300 MPa) and different weight percentages of $Bi_2O_3$ (5 wt%, 10 wt%, 15 wt%) in Zirconium dioxide and Zirconium hydroxide systems were employed. The resultant density and hardness values were tabulated to assist in further material property design. Also, phase composition and microstructural analysis were performed.

## II. METHODS

### A. Spark Plasma Sintering

Zirconium (IV) hydroxide (97 %, powder, density-3.25 $g/cm^3$ ) from Sigma-Aldrich, Zirconium (IV) oxide (99%, powder density-5.89 $g/cm^3$) , Bismuth (III) oxide (99.8%, powder, size: 90-210 nm, density-8.937 $g/cm^3$ ), both from Aldrich Chemistry were used as the starting powders. Spark plasma sintering of these ceramic powders was performed using the FCT Systeme Gmbh SPS machine. The powder systems (a) Pure Zirconium hydroxide, (b) pure Zirconium oxide, (c) Zirconium hydroxide - $Bi_2O_3$ sintering aid, and (d) Zirconium oxide - $Bi_2O_3$ sintering aid were all sintered under vacuum are hereafter referred to as (a), (b), (c), and (d) respectively. The powder systems (a) and (b) were ground using mortar and pestle before filling them in the die for sintering, whereas powder systems (c) and (d) were prepared by mixing them in the required ratio in a planetary ball mill with 2-propanol as a disperser. Since these powders are insoluble in alkalis and water, no change in the composition due to the wet mixing is expected. The 'ball mass: powder mass' ratio was always maintained as 4:1. They were wet-mixed for 3 hours at a 300-rpm rotation speed, and the slurry was later dried overnight in the fume hood under atomosphere. These powders were also ground using mortar and pestle before filling them in the die prior to sintering.

Dies made of either (i) Graphite (maximum mechanical loading-50Mpa) or (ii) Stainless steel (maximum mechanical loading-150Mpa), or (iii) Tungsten carbide-Cobalt (maximum mechanical loading-300Mpa) were used depending on the dwell piston pressure of the sintering cycle. The inner surface of the hollow cylindrical die was covered with thin Graphite foils to ensure uniform electric current conduction throughout the die. The powders were then weighed and filled in the die with thin Graphite discs at both end surfaces for uniform electrical conduction. The die assembly was then pressed under a hydraulic press under 1 ton force (10 kN) to prepare the cylindrical green body. The die was placed in the SPS machine and sintered with a uniform temperature ramp of 50 °C/min to the dwell temperature. The dwell temperature is kept constant for 10 minutes before free cooling the system back to ambient temperature. The mechanical loadings were also ramped up simultaneously with the temperature to reach the dwell piston pressure, kept constant until the end of the cycle, even while cooling the system. The whole process was performed under vacuum. The same cycle was followed for all the SPS experiments, as depicted in fig. 1

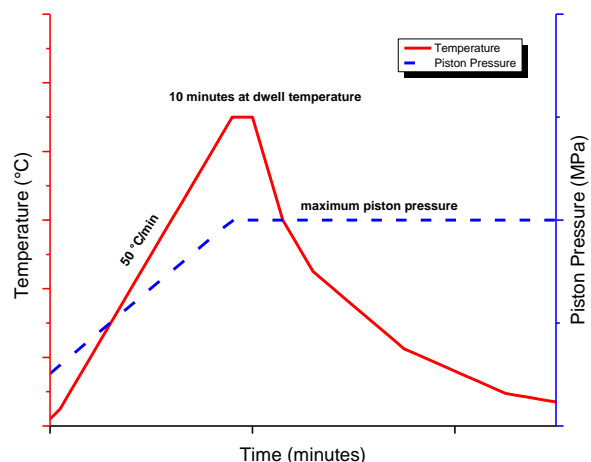

Fig. 1 SPS Pressure-Temperature profile followed for all the experiments

### B. Material Characterisation

Zirconium hydroxide powders were analyzed for their thermal behavior using Thermogravimetric analysis and Differential Scanning Calorimetry (DSC). These tests were also helpful to confirm the conversion to Zirconium oxide. For the TGA, the powders were heated from ambient temperature to 900°C with a heating rate of 50°C/min to mimic the sintering cycle used in the experiments. For the DSC, the powders were heated from ambient temperature up to 600°C and back to ambient temperature with a heating and cooling rate of 5°C/min to closely measure heat flow at each temperature step.

The microstructural analysis of as-sintered pellets after the SPS cycle was analyzed performed with a JEOL JSM7000F Scanning Electron Microscope (SEM). The extent of sintering in the sample was examined by observing the microstructure in comparison to relative density data. Energy Dispersive Spectroscopy (EDS) was performed on the same samples at random spots to verify the conversion to Zirconium oxide via the measured stoichiometry of Zr:O.

X-ray Diffraction (XRD) spectroscopy using Rigaku MiniFlex 600 was performed to measure the phase composition, especially different phases of Zirconia. . Cu-α X-rays were used with a set diffraction angle range of 2Θ = 10°-80°.

### C. Density and Hardness measurements

Geometrical dimensions of the as-sintered pellets were used to calculate the density of the pellets instead of Archimedes' method since the pellets have large porosity and are prone to property alteration due to solvent absorption. Relative densities were measured with respect to the density of tetragonal Zirconia, i.e., 6.10 g [21], on which the aerospace application of the ceramic is based. Since the density of tetragonal $ZrO_2$ is the highest among all its different phase forms while the sintered sample often contain other $ZrO_2$ phases such as monoclinic (5.68 g/cm$^3$) and cubic phase (6.09 g/cm$^3$), the calculated density ( $\frac{\text{Density of pellet } (g/cm^3)}{6.10 \, g/cm^3} * 100$ .) gives a conservative lowest estimation. It was calculated using the formula

The hardness of the as-sintered pellets was measured using the Leitz Durimet micro Vickers hardness testing machine. A counterweight of 500P (0.5Kg=4.905N) for around 15 seconds was applied to indent the sample surface. A magnified image of the indentation aided the measurement of the diagonals of the intender. The Vickers hardness number was calculated using the formula $HV = \frac{0.1891 * F}{d^2}$ where; HV is Vickers hardness (N/mm$^2$), F is the applied force = 4.905 N, and d is the average length of the diagonals (mm).

## III. RESULTS AND DISCUSSION

### A. Conversion to Zirconium oxide

The decomposition of Zr(OH)$_4$ to ZrO$_2$ is accompanied by dehydration from the system.

$$Zr(OH)_4 \rightarrow ZrO_2 + 2H_2O \uparrow \quad (1)$$

Since water is the by-product that will be released from the hydroxide powders during the reaction, the ideal mass loss on the complete conversion of Zr(OH)$_4$ to ZrO$_2$ is calculated as follows,

The molecular mass of Zr(OH)4 = 159.22g/mol
The molecular mass of ZrO2 = 123.22g/mol
The molecular mass of H2O = 18g/mol

$$Zr(OH)_4 \rightarrow ZrO_2 + 2H_2O \uparrow$$

The ideal mass loss % for the complete reaction would then be, 36.00/159.22*100=22.61%.

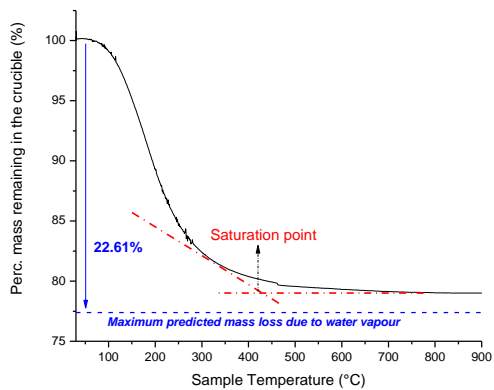

Fig. 2 Thermogravimetric analysis of the pure Zr(OH)$_4$ powders

It can also be seen from the TGA analysis with 50 °C/min heating ramp up to 900 °C shown in Fig. 2 that there is a slope change at around 200 °C, moreover, at beyond 400 °C, the conversion saturates. The mass loss at saturation (21.3 %) deviatesreasonably small (5.8%) from the calculated theoretical value (22.61%), signifying that the conversion to oxides is indeed accompanied by the dehydration of the precursor hydroxide powders.

The DSC curve from Fig. 3 shows an endothermic heat change around between 100 °C and 200 °C. The peak at 185 °C can be attributed to the initial dehydration of the hydroxide powders after the boiling point of water is reached. The curve undergoes a significant shift at 403 °C, where it reaches an exothermic peak corresponding to the crystallization of amorphous to crystalline zirconium oxide, which again confirms the Zr(OH)$_4$ to ZrO$_2$ conversion.

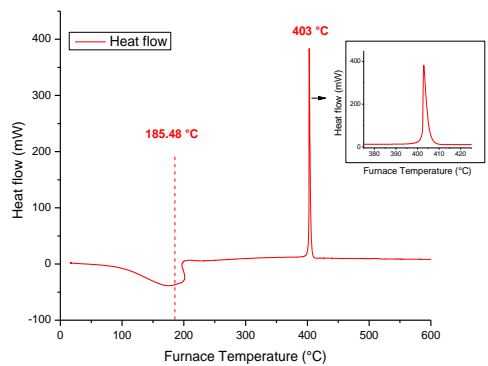

Fig. 3 Differential scanning calorimetry of the pure Zr(OH)$_4$ powders

Mass spectrometry embedded with the DSC machine measured the outlet gas composition from the pure Zr(OH)$_4$ system. In Fig. 4, it is shown that water liberation from the system was detected beginning from 100°C, which reached the maximum at 185 °C and finished at about 200 °C. This agrees well with the DSC measurement. At temperatures higher than 200 °C, water detection slowly decreased to $2.0 * 10^{-8}$ g (10mg - total sample weight), which corresponds well to the slope change in the TGA analysis. Another small peak of water detection occurs at 405 ºC, which agrees well with DSC peak at 403 ºC and could be explained by different amount of water crystallization in amorphous and crystalline ZrO$_2$. The water detection approaches towards zero at temperatures above 405 ºC and during cooling stage as well, which supports the saturation observation found in TGA analysis (Fig. 2).

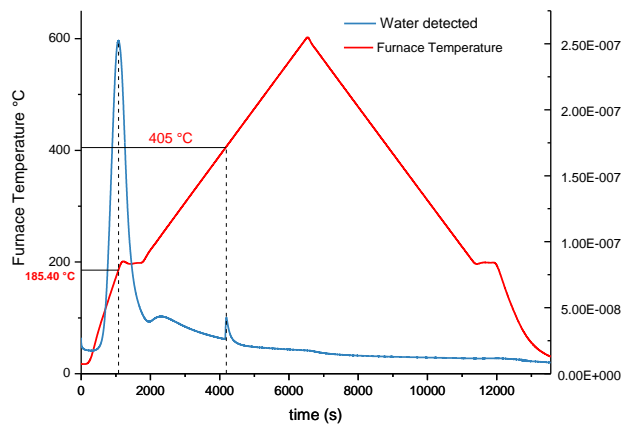

Fig. 4 Mass spectrometry associated with the temperature behaviour of Zr(OH)$_4$

### B. Influence of cold sintering due to Zr(OH)$_4$ to ZrO$_2$ conversion on densification of ZrO$_2$

It can be seen from Fig. 5 that for pure ZrO$_2$ systems, temperature and pressure do not influence sintering as evidentfrom their relative densities, which remain in the similar range even with the increase in temperature (from 400 ºC to 900 ºC) and pressure (from 50 MPa to 300 MPa). Whereas, the pure Zr(OH)$_4$ system samples exhibit a significant enhancement in the relative density values with both the increasing temperature and the increasing piston pressure. A combination of the effect of cold sintering process and high pressure assisted locking of water vapor within the sintering body and rearrangement of the powder particles is likely the reason for this distinctive contrast in sintering behavior between $Zr(OH)_4$ and $ZrO_2$..

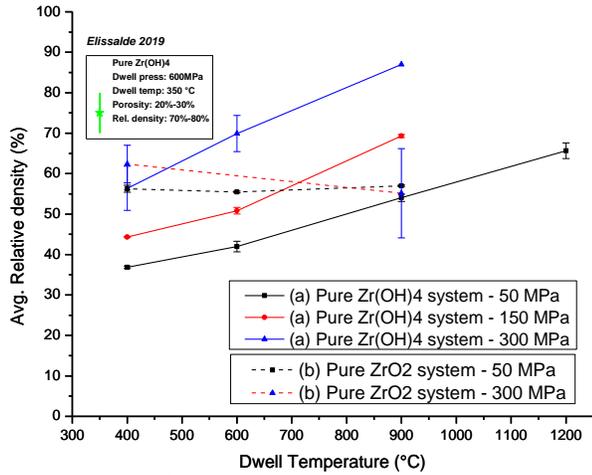

Fig. 5 Influence of sintering temperature and pressure on relative density for cold sintered $Zr(OH)_4$ and $ZrO_2$

*C. Influence of sintering aid- $Bi_2O_3$ on densification of $Zr(OH)_4$*

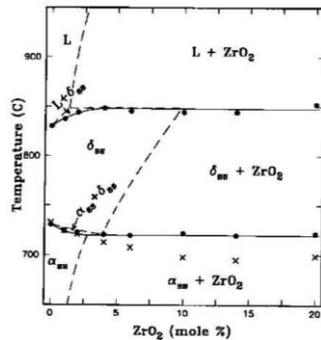

Fig. 6 Bi2O3-ZrO2 phase diagram [19]

The phase diagram [19] of the $Bi_2O_3$-$ZrO_2$ system, as shown in Fig. 6, denotes that at small concentrations of $ZrO_2$, Bismuth oxide exists as either α-$Bi_2O_3$ (monoclinic) or δ-$Bi_2O_3$ (cubic) or in liquid phase [22] depending on the temperature. In the sintering temperature range we are exploring (400 °C to 900 °C), these is small solubility of $ZrO_2$ in $Bi_2O_3$, which fulfills one of the most important prerequisites of liquid phase sintering. The effect of these $Bi_2O_3$ phases on the sintering of Zirconia is explored in this section.

Fig. 7 shows the influence of sintering aid – $Bi_2O_3$ on the relative densities of the corresponding samples. It is observed that at the same piston pressure of 50 MPa and dwell temperature of 900 °C, the addition of $Bi_2O_3$ increased the relative density by 25–30% when the composition was, either equal to more than 10 wt% $Bi_2O_3$. However, at 1200 °C, the 5wt% (c) sample increased the relative density by almost 35%, whereas the 10wt% and 15wt% (c) samples increased the relative densities by 15% and 25% respectively. The 5wt% (c) samples show higher densities than the 10wt% (c) and 15wt% (c) samples The lattice changes associated with phase change of the monoclinic Zirconia into tetragonal Zirconia around 1200 °C, influencing the solubility of bismuth oxide in $ZrO_2$, can be considered as a reason for this phenomenon Further experiments of solubility study are needed to clarity, given no existing phase diagram data on the $ZrO_2$-rich end.

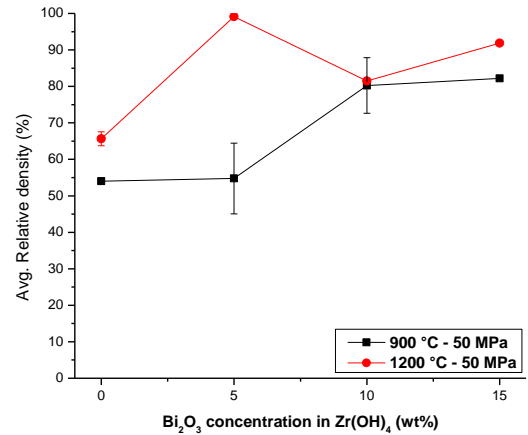

Fig. 7 Influence of sintering aid-$Bi_2O_3$ on the relative density

*D. Influence of sintering conditions (temperature and pressure) and sintering aid- $Bi_2O_3$ on hardness*

The surface hardness is another measure to indicate the extent of sintering in bulk. Fig. 8 shows the hardness dependency upon temperature, pressure, and addition of $Bi_2O_3$. For system (a) samples, the hardness increases with increasing temperature and piston pressure in line with the increasing densities of the samples. The decrease of porosities associated with increased densities is responsible for such behavior. Pure $Zr(OH)_4$ sample sintered at 300 MPa - 600 °C dwell conditions had a similar hardness value as the one from the samples tested by Elissalde et al. [13] (3.805 GPa) with sintering dwell condition 350 ºC – 600 MPa. Also, the hardness of the 10wt% $Bi_2O_3$-$Zr(OH)_4$ sample (6.59GPa) sintered at 900 °C - 50 MPa piston pressures surpassed this value from the literature, showing prospects of tuning mechanical properties by $Bi_2O_3$ doping.

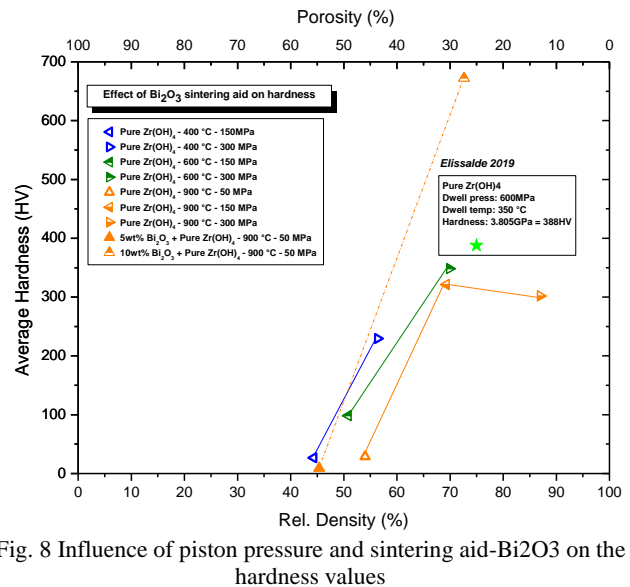

Fig. 8 Influence of piston pressure and sintering aid-Bi2O3 on the hardness values

*E. Phase changes*

Literature [23] found that the Zirconia's tetragonal phase reflections are observed at around 30.25°, and the monoclinic reflections at around 28.20° and 31.47°.

Fig. 9 shows the XRD spectrum comparison between system (a) sample and system (c) sample sintered at the same dwell conditions of 50 MPa and 900 ºC dwell temperatures. The sintering aid $Bi_2O_3$ in the starting powders also enhanced the tetragonal phases in the sintered Zirconia system. The 10wt% $Bi_2O_3$ system had a higher tetragonal phase stabilization effect than the 5wt% $Bi_2O_3$ system

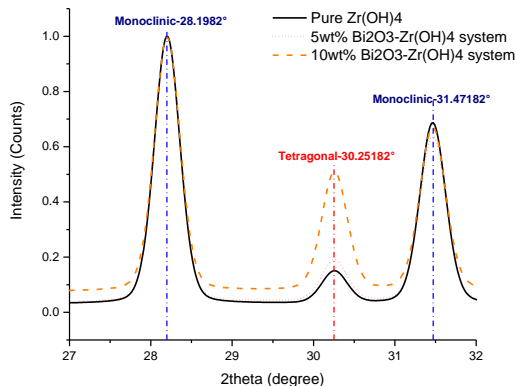

Fig. 9 XRD plots of sintered pure $Zr(OH)_4$, sintered 5wt% and 10wt% $Bi_2O_3$-$Zr(OH)_4$ system at 900 °C, 50 MPa

## IV. CONCLUSION

The in-situ liberation of water vapour from the precursors Zirconium hydroxide powders along with the increased piston pressure enhanced the relative densities and hardness of the as-sintered samples. The 10wt% $Bi_2O_3$-$Zr(OH)_4$ system sintered at 900 ºC - 50MPa dwell conditions had the highest harness of value of 6.6 GPa, whereas the 5wt% $Bi_2O_3$-$Zr(OH)_4$ system had the highest density among all the samples reaching around 99% relative density. The 10wt% $Bi_2O_3$-$Zr(OH)_4$ systems also stabilized the tetragonal phase of Zirconia in the sintered samples better than the 5wt% $Bi_2O_3$-$Zr(OH)_4$ systems.